White Paper

# A Blockchain-based Carbon Credit Ecosystem

Dr. Soheil Saraji
Dr. Mike Borowczak

Assistant Professor, Department of Petroleum Engineering, University of Wyoming
Assistant Professor, Department of Computer Science, University of Wyoming

**Abstract.** Climate change and global warming are the significant challenges of the new century. A viable solution to mitigate greenhouse gas emissions is via a globally incentivized market mechanism proposed in the *Kyoto protocol*. In this view, the carbon dioxide (or other greenhouse gases) emission is considered a commodity, forming a carbon trading system. There have been attempts in developing this idea in the past decade with limited success. The main challenges of current systems are fragmented implementations, lack of transparency leading to over-crediting and double-spending, and substantial transaction costs that transfer wealth to brokers and agents. We aim to create a Carbon Credit Ecosystem using smart contracts that operate in conjunction with blockchain technology in order to bring more transparency, accessibility, liquidity, and standardization to carbon markets. This ecosystem includes a tokenization mechanism to securely digitize carbon credits with clear minting and burning protocols, a transparent mechanism for distribution of tokens, a free automated market maker for trading the carbon tokens, and mechanisms to engage all stakeholders, including the energy industry, project verifiers, liquidity providers, NGOs, concerned citizens, and governments. This approach could be used in a variety of other credit/trading systems.

## 1. Introduction

*1.1.* **Carbon Emission.** The global Greenhouse Gas (GHG) emissions trend has increased since the beginning of the 21st century compared to the three previous decades, mainly due to the increase in $CO_2$ emissions from the emerging economies. As a result, the atmospheric concentrations of greenhouse gases substantially increased, worsening the natural greenhouse effect, which negatively affects the life on the Earth. $CO_2$ emissions, the main gas responsible for global warming, are still increasing at the world level despite climate change mitigation agreements[1]. The 2018 GHG emissions amounted to 55.6 $GtCO_2$ eq. Present GHG emissions are about 57% higher than in 1990 and 43% higher than in 2000 [2]. **Figure 1** shows the increasing trend of main GHG emissions based on sector.

Under Kyoto Protocol (1997), seven greenhouse gases are considered as the significant contributors to global warming: Carbon dioxide ($CO_2$), Methane ($CH_4$), Nitrous oxide ($N_2O$), Hydrofluorocarbons (HFCs), Perfluorocarbons (PFCs), Sulphur hexafluoride ($SF_6$), Nitrogen trifluoride ($NF_3$)[3]. $CO_2$ emissions from fossil fuels are the largest source of global GHG emissions, with a share of about 72%, followed by $CH_4$ (19%), $N_2O$ (6%), and F-gases (3%) [2]. The direct drivers of $CO_2$ are the combustion of coal, oil, and natural gas, representing 89% of global $CO_2$ emissions, with respective shares of 39%, 31%, and 18% [2]. For $CH_4$, there are three large sources: agriculture, fossil fuel production, and waste/wastewater. Together, fossil fuel production and transmission account for a third of global methane emissions [2].

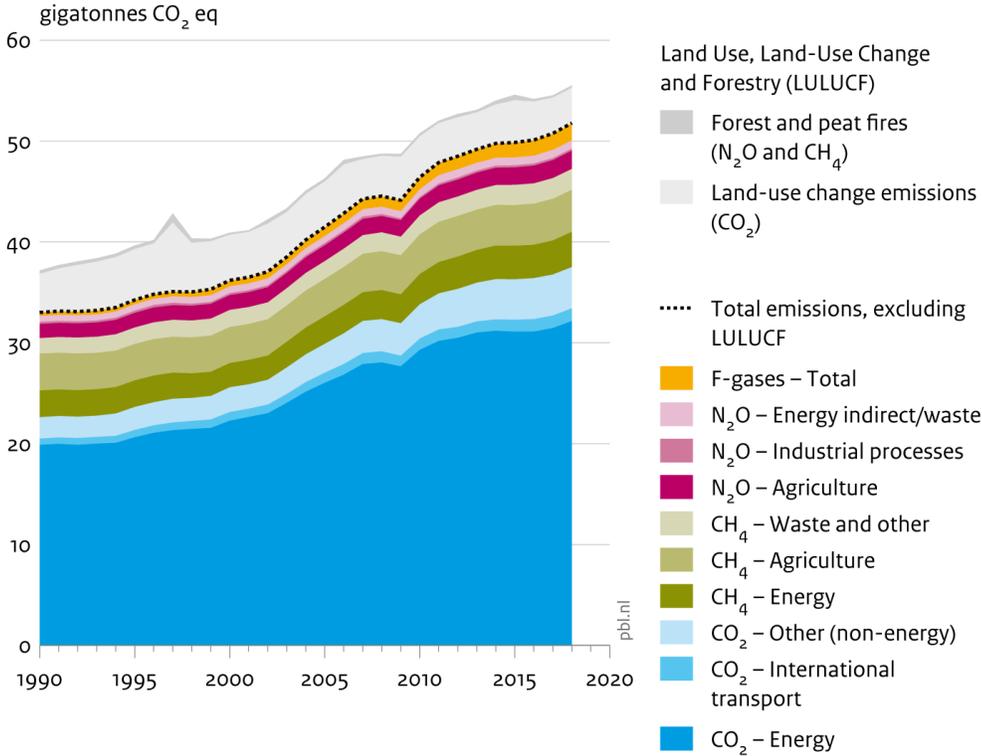

**Figure 1.** Global GHG emission per type of gas and emission source [3].

***1.2. Carbon Credits.*** In December 2015, the *Paris Agreement* brought all nations into a common cause to undertake ambitious efforts to combat climate change. It required all parties to agree to put forward their best efforts through 'nationally determined contributions'[1]. At the national level, policymakers have three options to reduce greenhouse gas emissions: (1) set a specific limit that a company cannot exceed, (2) introduce a carbon tax where the company pays for the amount of $CO_2$ they produce, (3) implement an emission trading scheme – to create a carbon market. The last solution has been gaining traction recently because of the positive encouragement of clean producers of energy and incentivizing the Fossil fuel industry to become more efficient and gradually reduce their emissions. This led to the creation of carbon credits. A carbon credit is a tradable permit or certificate that gives the right to emit one ton of carbon dioxide or an equivalent of another greenhouse gas. There are currently two types of carbon credits: (1) voluntary emissions reduction (VER): a carbon offset exchanged in the over-the-counter or voluntary market for credits, (2) certified emissions reduction (CER): emission units (or credits) created through a regulatory framework with the purpose of offsetting a project's emissions. The main difference between the two is that there is a third-party certifying body that regulates the CER as opposed to the VER.

*1.3.* *Carbon Credit Markets.* *Kyoto protocol* regards the market mechanism as a way to solve the greenhouse gas emission reduction problem. In this view, carbon dioxide emission is considered a commodity, forming a carbon trading system. According to the United Nations, carbon offsetting is particularly crucial for meeting the *Paris Climate Agreement*'s goal. There have been efforts to mitigate this problem using a carbon tax or credit. Carbon offsetting allows companies and individuals to reduce carbon emissions by purchasing carbon credits from carbon reduction projects. These projects include planting new trees, avoiding deforestation, investment in renewable energy, carbon capture, and sequestration projects. The voluntary carbon market co-exists with the compliance carbon markets, driven by regulatory caps on greenhouse gas emissions, and operates at a significantly larger scale. The New York Times reported that greenhouse gas (GHG) emissions trading held the potential to become the world's largest commodity market [4]. According to the World Bank's tally, the carbon emission trading market reached about 10.9 billion USD in 2005 when the Kyoto Protocol took effect and continued to grow at an annual rate of 108%, growing to 143.7 billion USD in 2009 [5]. Recently, carbon offset and emissions trading volume grew 34% and peaked at 215 billion USD in 2019, with over 80% of the total volume accounted for by the EU's Emissions Trading System (ET ETS) [6].

*1.3.1.* *How it works.* The number of permits in the market is capped. At the beginning of a trading phase, emission permits are either allocated to businesses for free or have to be bought at auction. The number of available permits decreases over time, putting pressure on the participating companies to invest in cleaner production options and reduce their $CO_2$ outputs. In the long run, this fuels innovation and drives down the price of new technologies. On the other hand, carbon reduction projects will be awarded carbon offsetting credits for the removal of GHGs from the atmosphere. These projects include planting new trees, avoiding deforestation, investment in renewable energy, carbon capture, carbon sequestration projects in the deep saline aquifers or depleted oil and gas reservoirs, etc. Carbon emitters could buy carbon credits from the carbon reduction projects to temporarily increase their emission amount if they fail to achieve their set emission caps. For instance, assume the regulatory body allots company A 100 units of GHG emissions. This company ends up emitting 120 units of GHG by the end of the year. Landowner B preserves forest and receives 50 carbon offset credits. Company A buys 20 offset credits from landowner B at the market price to increase its emission cap.

*1.3.2.* *Current Challenges.* The growth of the global GHG emissions markets has caused the emergence of an array of global and regional credits, markets, and trading mechanisms [7,8]. The prominent examples are the European Emission Trading Scheme (EU ETS), New Zealand Emissions Trading (EZ ETS), Midwestern Greenhouse Gas Accord (MGA), and California's cap-and-trade program [8]. This naturally resulted in the fragmented implementation of the cap-and-trade-scheme and lack of cross-market exchange of value. These programs have also been criticized for over-crediting, unclear life-cycle of issued carbon credits, and promoting double-spending due to the lack of transparency and inter-regional communications [9]. The other significant problem with the current carbon trading markets is substantial transaction costs that transfer wealth to brokers and agents. Traders and brokers often get a commission of 3 to 8 percent of the value of the credit (with an industry average of 5 percent) [7]. Furthermore, these schemes rely on third-party verifiers to check claims and often are paid by project developers, meaning they have an incentive to approve all clean projects they investigate [7]. The confluence of these factors may explain why global (and regional) carbon markets have failed to make a meaningful difference in mitigating greenhouse gas emissions so far.

## 2. Blockchain Technology

*2.1. Background.* Blockchain technology enables an immutable ledger that is distributed, validated, and agreed upon by the parties that participate in its operation. The foundations for Blockchain are rooted in the ability to achieve a consensus of distributed parties, a problem that has been solved in one way or another since the early 1980s (by Lamport et al.)[10]. Redundant, consensus-backed systems have existed in mission-critical spaces for decades, but technology limitations meant that the implementation overhead made these systems impractical and cost-ineffective until the late 2000s. The collapse of the financial markets exposed critical trust-based issues related to double-spending, which formed the impetus for Satoshi's seminal work to leverage peer-to-peer distributed systems to generate computational proof of chronologically ordered transactions [11]. With a way to order and link transactions, Vitalik proposed the formation of decentralized autonomous organizations (DAOs) controlled by automated pre-programmed rules [12]. These automated smart contracts could contain assets, encode, and execute bylaws of an entire organization. The automated transactions from these contracts could be collected at some regular intervals and placed in a block, and chained together.

Blockchain technology finds its name rooted in the structure of the data it captures. The fundamental component of a blockchain is a block of new transactions as well as information about the previous block. The ability to link the new block to the previous block in proof of work systems requires a cryptographic puzzle that is computationally hard to solve by easy to validate its solution. Once linked, information about the newly added block, which recalls also includes information from the previous block, is passed along for inclusion in the next block of transactions, and the process repeats. The inclusion of information about a block's "parent block" or proceeding block creates a structure that would require immense computational power to subvert.

*2.2. Benefits of Blockchain for Carbon Trading.* Blockchain technology provides a safe and reliable, efficient and convenient, open, and inclusive platform that is uniquely suited for implementing Carbon Credit Markets. The immutable cryptographically-secured distributed ledger on the Blockchain allows for reliable issuance and tracking of carbon credits. Public blockchains are easily accessible to small and medium-sized enterprises, reducing the entry threshold for the carbon trading market. Furthermore, the information provided by companies is transparent and accessible to everyone. Recently, free automated market makers (AMMs) have been developed on blockchains allowing for the trading of digitized assets directly on the Blockchain without intermediary and minimal algorithmic fees. They provide the infrastructure required to create a digital carbon credit ecosystem and engage all the stakeholders.

## 3. A Blockchain-based Carbon Credit Ecosystem

*3.1.* *Vision.* We aim to create a Carbon Credit Ecosystem on Blockchain to bring more liquidity, transparency, accessibility, and standardization to carbon markets. This ecosystem includes all stakeholders, a tokenization mechanism with clear minting and burning protocols, a transparent distribution of tokens, and an AMM for trading these carbon tokens.

*3.2.* *Work Plan.* In this project, different stakeholders involved are "Generators" of carbon credit (i.e., wind farms, tree-planting operations, $CO_2$ sequestration projects, etc.) and "Consumers" of carbon credit (i.e., carbon emitters or polluters of any kind such as the energy industry) as well as other stakeholders such as regulators, concerned citizens, and validators. "Validators" are an essential part of this ecosystem. They are accredited, globally distributed, technically competent consultants incentivized to parameterize appropriately and onboard projects to an open architecture marketplace that matches interested parties generating and retiring carbon credits.

We will transfer carbon credits to the Blockchain by converting them into digital tokens distributed to carbon credit generators after properly validating their projects. Buyers and sellers of carbon credit will use a decentralized exchange platform on Blockchain to trade Carbon credits. The price will be determined by market dynamics driven by supply and demand. The Carbon Tokens would be retired via a "buy and burn" model by sending the given Carbon Tokens to a smart contract or defined blockchain address whose private key is not known by any party and can be visible to the collective of validators as well as regulators or other stakeholders. The companies and individuals who successfully burn their Carbon Tokens will be issues non-fungible tokens as a carbon removal certificate.

*3.3.* *Smart Contracts.* This project's minimal requirement involves four (4) smart contracts interacting with three (3) stakeholders and liquidity providers. A simplified flow-chart is illustrated in **Figure 2**, and the design requirements for each smart contract are listed below.

*3.3.1. Smart Contract 1.* A registry system on the Blockchain to record the essential information for the following stakeholders: (a) *Verifiers:* They validate carbon credits from credit holders. They also verify that carbon tokens burnt are equivalent to burning a proportionate amount of emissions. (b) *Credit-holders:* They are organizations that already hold carbon credits in the emissions trading ecosystem. (c) *Customers:* They are individuals and companies interested in offsetting their carbon footprint by buying carbon credits and burning the carbon token.

*3.3.2. Smart Contract 2.* A smart contract to mint digital tokens called carbon tokens based on studied and approved carbon credits through a series of functions: (a) Approve credits entered by the credit-holders, which is certified by the verifiers, (b) Mint the Carbon token, (c) Make the Carbon token transferrable and burnable, (d) Burn offset carbon tokens, (e) Mint non-fungible tokens as a badge of successful burning carbon tokens, representing offsetting carbon emissions

*3.3.3. Smart Contract 3.* A smart contract with a multi-signature allows verifiers to verify the minting and burn the carbon tokens. This contract will require approval by at least 70% of the verifiers to be automatically executed.

*3.3.4. Smart Contract 4:* An automated market maker (AMM) smart contract that allows: (a) automated trading of Carbon tokens with digital money (e.g., stable coins or future central bank digital currencies), (b) provides incentives for liquidity providers (LPs) by charging a transaction fee (example: 0.3%) and distributes among LPs, (c) provides a dynamic price discovery for the Carbon tokens in a free market.

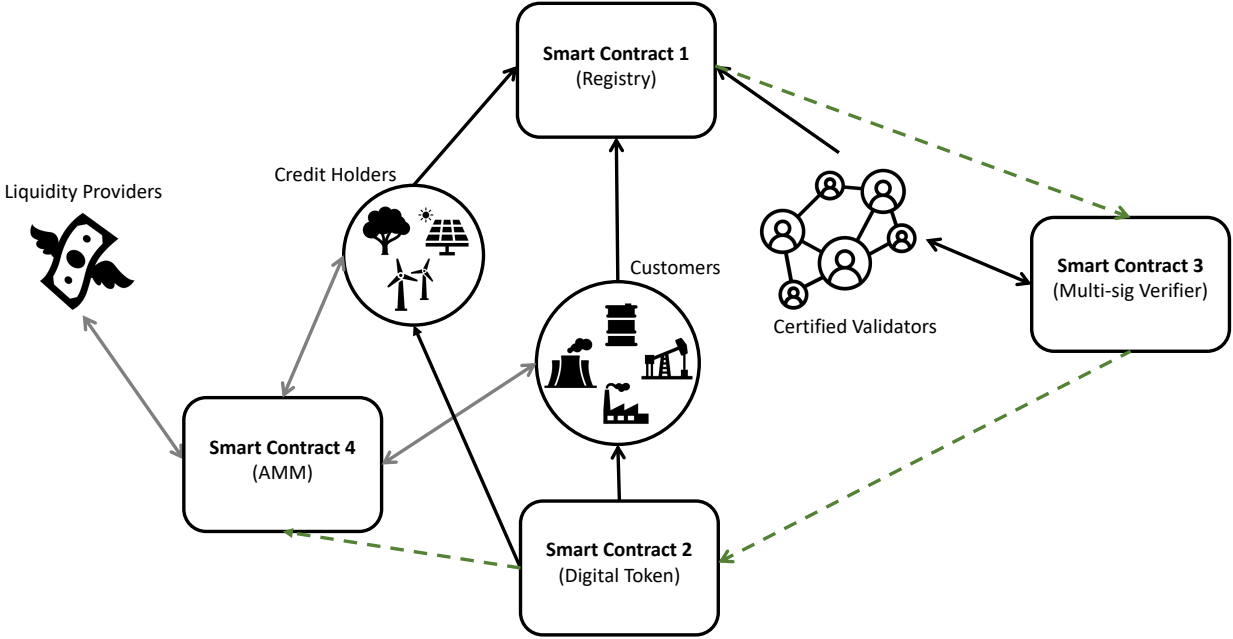

**Figure 2.** Flow diagram showing the proposed Carbon credit ecosystem on a blockchain.

*3.4.  Previous Work.* The PI attended the Ethglobal Hackathon in Fall 2020. There, he led a group of international developers to create the first version of this project on the Ethereum blockchain. The project was called Carbon Project and involved three smart contracts deployed on an Etheremu testnet. More information about this project is provided below.

*3.4.1. The public URL for the Carbon Project in Ethglobal Hackathon*

hack.ethglobal.co/showcase/carbon-credit-project-recye9DJnVvSE8vPG

*3.4.2. Github repository including smart contracts for Carbon Project*

https://github.com/CarbonCreditProject/Carbon-Project.git

**References:**
1. Crippa, M., Guizzardi, D., Muntean, M., Schaaf, E., Solazzo, E., Monforti-Ferrario, F., Olivier, J.G.J., Vignati, E., Fossil $CO_2$ emissions of all world countries - 2020 Report, EUR 30358 EN, Publications Office of the European Union, Luxembourg, 2020, ISBN 978-92-76-21515-8, doi:10.2760/143674, JRC121460.
2. Olivier, J.G.J., and Peters, J.A.H.W., (2019), Trends in global CO2 and total greenhouse gas emissions: 2019 report. PBL Netherlands Environmental Assessment Agency, The Hague.
3. European Commission, Joint Research Centre (EC-JRC)/Netherlands Environmental Assessment Agency (PBL). Emissions Database for Global Atmospheric Research (EDGAR), release EDGAR v5.0 (1970 - 2015) of November 2019.
4. Kanter, J., (2007), Banks Urging US to Adopt Trading of Emissions, New York Times, 26 September 2007.
5. Kim, S.-K., and Huh, J.-H., (2020), Blockchain of Carbon Trading for UN Sustainable Development Goals, Sustainability, 12, 4021; doi:10.3390/su12104021
6. Nordeng, A., et al., (2020), Refinitiv Carbon Market Survey, May 2020.
7. Sovacool, B.K., (2011), Four Problems with Global Carbon Markets: A Critical Review, Energy & Environment · Vol. 22, No. 6, 201.
8. Gale, K., Langer, D., Waterman, R. (2010), Global Carbon Credit Markets – Issues and opportunities, pfi market intelligence/Financing clean energy, Section 4, chapter 39.
9. Haya, B., Cullenward, D., Strong, A. L., Grubert, E., Heilmayr, R., Sivas D. A., Wara, & M. (2020) Managing uncertainty in carbon offsets: insights from California's standardized approach, Climate Policy, 20:9, 1112-1126, DOI: 10.1080/14693062.2020.1781035.
10. Lamport, L.; Shostak, R.; Pease, M. (1982). The Byzantine Generals Problem. ACM Transactions on Programming Languages and Systems. 4 (3): 387–389. doi:10.1145/357172.357176.
11. Nakamoto, S. (2008), Bitcoin: A Peer-to-Peer Electronic Cash System, Bitcoin White Paper.
12. Buterin, V. (2016), A Next Generation Smart Contract and Decentralized Application Platform, Ethereum White Paper.